\documentclass[aps,prl,twocolumn,showpacs,amssymb, footinbib]{revtex4-1}
\usepackage{amsfonts,amsmath,amssymb,amsthm,bbm,hyperref}

\usepackage{color,psfrag}
\usepackage[dvips]{graphicx}

\usepackage[normalem]{ulem}

\newcommand{\vev}[1]{\left\langle #1 \right\rangle}

\newcommand{\m}{\mbox{}}

\newcommand{\be}{\begin{equation}}
\newcommand{\ee}{\end{equation}}
\newcommand{\ba}{\begin{eqnarray}}
\newcommand{\ea}{\end{eqnarray}}

\begin{document}

\title{Loop quantum gravity without the Hamiltonian constraint}

\author{N. Bodendorfer$^{1,2}$}

\author{A. Stottmeister$^{1}$}

\author{A. Thurn$^1$}

\affiliation{$^1$Inst. for Theoretical Physics III, FAU Erlangen -- N\"urnberg, Staudtstr. 7, 91058 Erlangen, Germany\\}

\affiliation{$^2$ Institute for Gravitation and the Cosmos \& Physics Department, Penn State, University Park, PA 16802, U.S.A.}

\date{{\small\sf \today}}

\begin{abstract}

We show that under certain technical assumptions, including the existence of a constant mean curvature (CMC) slice and strict positivity of the scalar field, general relativity conformally coupled to a scalar field can be quantised on a partially reduced phase space, meaning reduced only with respect to the Hamiltonian constraint and a proper gauge fixing. More precisely, we introduce, in close analogy to shape dynamics, the generator of a local conformal transformation acting on both, the metric and the scalar field, which coincides with the CMC gauge condition. A new metric, which is invariant under this transformation, is constructed and used to define connection variables which can be quantised by standard loop quantum gravity methods. Since this connection is invariant under the local conformal transformation, the generator of which is shown to be a good gauge fixing for the Hamiltonian constraint, the Dirac bracket associated with implementing these constraints coincides with the Poisson bracket for the connection. Thus, the well developed kinematical quantisation techniques for loop quantum gravity are available, while the Hamiltonian constraint has been solved (more precisely, gauge fixed) classically. The physical interpretation of this system is that of general relativity on a fixed spatial CMC slice, the associated ``time'' of which is given by the constant mean curvature. While it is hard to address dynamical problems in this framework (due to the complicated ``time'' function), it seems, due to good accessibility properties of the CMC gauge, to be well suited for problems such as the computation of black hole entropy, where actual physical states can be counted and the dynamics is only of indirect importance. The corresponding calculation yields the surprising result that the usual prescription of fixing the Barbero-Immirzi parameter $\beta$ to a constant value in order to obtain the well-known formula $S = a(\Phi) A/(4G)$ does not work for the black holes under consideration, while a recently proposed prescription involving an analytic continuation of $\beta$ to the case of a self-dual space-time connection yields the correct result.
Also, the interpretation of the geometric operators gets an interesting twist, which exemplifies the deep relationship between observables and the choice of a time function and has consequences for loop quantum cosmology.

\end{abstract}


\maketitle

A reduced phase space quantisation of a given theory is generally very problematic due to the complexity of the representation problem resulting from a non-trivial Dirac bracket. When quantising a given classical theory, it is often more practical to perform a Dirac-type quantisation \cite{DiracLecturesOnQuantum} and to represent the constraints of the classical theory on a kinematical Hilbert space, as for example done in loop quantum gravity \cite{RovelliQuantumGravity, ThiemannModernCanonicalQuantum}. On the other hand, the quantum equations are generally hard to solve and new technical problems, mostly of functional analytic nature, arise. 

Concerning general relativity, the Dirac-type quantisation has been performed in the context of loop quantum gravity (LQG), as well as quantisations based on deparametrisation with respect to matter fields \cite{GieselAQG4, DomagalaGravityQuantizedLoop, HusainTimeAndA, GieselScalarMaterialReference} have been given. While the focus of the available deparametrised models has been on solving the problem of time by deriving a true Hamiltonian, there are situations, e.g. state counting in the derivation of the black hole entropy, where the dynamics are not relevant, but only access to the physical Hilbert space is needed. For this, one would need a gauge fixing $\mathcal{D}$ (a time function) of the Hamiltonian constraint $\mathcal{H}$, i.e. $\{\mathcal{H}, \mathcal{D} \} = \text{invertible}$, which is accessible (at least for the specific situations under consideration) and leads to a manageable Dirac bracket. The interpretation of such a formulation would be to consider general relativity on a fixed spatial slice defined by the gauge fixing condition.

In this paper, we show how such a reduced phase space quantisation can be constructed for general relativity conformally coupled to a scalar field by using ideas from shape dynamics \cite{GomesEinsteinGravityAs}. First, we show that the generator of a local conformal symmetry (i.e., in what follows, a local rescaling of the canonical variables) is a good gauge fixing for the Hamiltonian constraint. Interestingly, the generator coincides with the constant mean curvature (CMC) gauge condition \cite{RendallConstantMeanCurvature}, thus being a purely geometric clock. A new metric, which is invariant under the local conformal transformation, is constructed as a compound object of the original metric and the scalar field. Due to this invariance, the Dirac bracket with respect to implementing both the Hamiltonian constraint and the generator of the conformal symmetry coincides with the Poisson bracket. Passing to Ashtekar-Barbero type connection variables, the Ashtekar-Isham-Lewandowski representation of loop quantum gravity can be employed. At the quantum level, we are left with the Gau{\ss} and spatial diffeomorphism constraints. The spatial diffeomorphism constraint poses the same difficulties as in standard LQG, e.g. the construction of spatially diffeomorphism invariant operators and the associated question of graph preservation need further research. 

As an application, we perform an entropy calculation for a family of topologically non-trivial black holes which can be treated by the proposed method. The state counting can thus be performed at the level of the physical Hilbert space and yields some surprising results.

Although access to the physical Hilbert space is also given in the models \cite{GieselAQG4, DomagalaGravityQuantizedLoop, HusainTimeAndA, GieselScalarMaterialReference}, it is unclear if their associated time functions can be good gauge fixings for the Hamiltonian constraint for the type of static black hole solutions under consideration in this paper. 
E.g. since $\{\mathcal{H}, \mathcal{D} \}$ would be linear in the momenta for scalar field or dust clocks $\mathcal{D}$, we have $\{\mathcal{H}, \mathcal{D} \}=0$ in static situations at least for the type of foliations (vanishing momentum of the scalar field, see the section on black holes) considered in this paper. Using other foliations, this objection would not hold, however, the CMC foliations considered here seem to be the most natural ones for static solutions.
Of course, in other situations, the time functions of \cite{GieselAQG4, DomagalaGravityQuantizedLoop, HusainTimeAndA, GieselScalarMaterialReference} might be better suited than the one presented here. Furthermore, we comment on using the variables derived in this paper for loop quantum cosmology.

\section*{Classical analysis}

The presentation of the calculations in this paper is concise, see our more comprehensive companion paper \cite{BSTII} for details. The action of general relativity conformally coupled to a scalar field is given by
\ba
\label{eq:CombinedAction}
	S = &&\frac{1}{\kappa} \int_{M} d^{D+1}X \sqrt{g} R^{(D+1)} a(\Phi)   \nonumber \\ &+& \frac{1}{2\lambda} \int_{M} d^{D+1}X \sqrt{g} g^{\mu \nu} (\nabla_{\mu} \Phi) (\nabla_{\nu} \Phi) \text{,}
\ea
where we defined
\be
	a(\Phi) := 1-\alpha \Phi^2 , ~~~\alpha := \frac{\kappa}{2 \lambda D} \frac{\Delta^{\Phi}}{\Delta^{g}},ÊÊ~~~ \Delta^{\Phi} = \frac{1-D}{4} \Delta^g \nonumber
\ee
and $D,\ \Delta^{g},\ \Delta^{\Phi}$ denote the spatial dimension and conformal weights of the metric and scalar field respectively. The dimension of $\lambda$ is chosen to coincide with the dimension of $\kappa$ which renders the field $\Phi$ dimensionless. The scalar field part of the action, i.e. $S - S_{\text{Einstein-Hilbert}}$, is invariant under the conformal transformation
\be
	 g_{\mu \nu}  \rightarrow \Omega^{\Delta^g} g_{\mu \nu}, ~~~~ \Phi \rightarrow \Omega^{\Delta^{\Phi}} \Phi  \text{.}
\ee
The $D+1$ split of this action gives
\be
S=\int_{\mathbb{R}} dt \int_{\sigma} d^Dx \left[P^{ab} \dot{q}_{ab} + \pi_{\Phi} \dot{\Phi} - N^a \mathcal{H}_a - N \mathcal{H} \right] \text{,}
\ee
where we have defined
\ba
	\pi_{\Phi} &:=&  -\frac{1}{\lambda} \sqrt{q} (\mathcal{L}_n\Phi) + \frac{4 \alpha}{\kappa} \sqrt{q} \Phi K \text{,} \nonumber \\ 
	P^{ab} &:=&  \frac{1}{\kappa} a(\Phi) \sqrt{q} \left(K^{ab} - q^{ab} K\right) \nonumber \\
		& &+  \frac{2\alpha}{\kappa} \sqrt{q} q^{ab} \Phi (\mathcal{L}_n \Phi) \text{,} \label{eq:ADMMomentum} \\
	P^{ab}_{\text{tf}} &:=& P^{ab} - \frac{1}{D} q^{ab} P^{cd} q_{cd} \text{,} \nonumber \\
	\mathcal{H}_a[N^a] &:=& \int_{\sigma} d^Dx \left[ P^{ab} (\mathcal{L}_N q)_{ab} + \pi_{\Phi} (\mathcal{L}_N \Phi) \right]\text{,} \\
	\mathcal{H}[N] &:=&  \int_{\sigma} d^Dx N \left[ \mathcal{H}_{\text{Grav}} + \mathcal{H}_{\Phi} - \frac{\kappa \; \mathcal{D}^2}{{\Delta^g}^2 D(D-1)\sqrt{q}}  \right]\text{,}\nonumber \\
	\kappa \mathcal{H}_{\text{Grav}} &:=&  \frac{\kappa^2}{\sqrt{q} a(\Phi)} P^{\text{tf}}_{ab} P_{\text{tf}}^{ab} - \sqrt{q} R^{(D)} \text{,} \nonumber 
\ea
\ba
		\kappa \mathcal{H}_{\Phi} &:=& \frac{\kappa}{2\lambda} \sqrt{q} \biggl[ - \frac{\lambda^2}{q} \pi_{\Phi}^2 - \frac{1}{D} q^{ab} (D_a \Phi) (D_b \Phi)\nonumber \\ && ~~~~~~~~~+  \frac{D-1}{D} \Phi D_a D^a \Phi + \frac{1}{D} \frac{\Delta^{\Phi}}{\Delta^g} R^{(D)} \Phi^2\biggr]\text{,} \nonumber \\
	\mathcal{D} &:=& \Delta^g P + \Delta^{\Phi} \pi_{\Phi} \Phi = \frac{\Delta^g (1-D) \sqrt{q}}{\kappa} K  \text{.} 
\ea
$n$ denotes the normal vector on the spatial slices, $\mathcal{L}$ the Lie derivative, $K_{ab}$ the extrinsic curvature, $D_a$ the covariant derivative compatible with the spatial metric $q_{ab}$, $K = K^{ab}q_{ab}$, $P = P^{ab}q_{ab}$, and $P^{ab}_{\text{tf}}$ denotes the trace free part of the spatial metric conjugate momentum $P^{ab}$. 
It is easy to see that $\mathcal{D}$ is the generator of local conformal transformations. 

The underlying idea of what follows originates in the work of Lichnerowicz \cite{LichnerowiczLIntegrationDes} and York \cite{YorkGravitationalDegreesOf}: Good initial data (satisfying $\mathcal{H}=0= \mathcal{H}_a$) for general relativity can be constructed from specific initial data (a spatial metric, a transversal trace free second rank tensor field and a constant value for the mean curvature) by performing a conformal rescaling of the fields with a scaling factor satisfying the Lichnerowicz-York equation. On the other hand, if, morally speaking, only conformal equivalence classes of initial data would be specified, one could perform a conformal transformation to initial data satisfying the Hamiltonian constraint without leaving the equivalence class, i.e. without changing the initial data. It therefore transpires that one should try to exchange the equation $\mathcal{H}=0$ for invariance under a local conformal rescaling. Parts of this idea have been implemented in shape dynamics \cite{GomesEinsteinGravityAs}, however, it was not possible so far to find a general solution to the conformal invariance condition which could also be quantised in a satisfactory way. 
In this paper, we take this last step by realising that a conformally coupled scalar field, as opposed to obstructions arising from other matter fields \cite{GomesCouplingShapeDynamics}, allows for a non-trivial conformal weight in the generator of the local conformal transformation, and thus for the construction of a conformally invariant metric. We remark that an earlier account of introducing a conformal symmetry in canonical quantum gravity has been given in \cite{WangConformalGeometrodynamicsTrue, WangTowardsConformalLoop}, however, to the best of our knowledge, the kernel of the quantised conformal constraint, which is a part of the constraint algebra, has not been studied so far.

The main result of this section is that the CMC gauge $\mathcal{D}=0$ is a good gauge fixing for the Hamiltonian constraint at least locally and restricting to spatial slices which allow for the $\mathcal{D}=0$ gauge. While we restrict to zero constant mean curvature in this paper, the general case is developed in our companion paper \cite{BSTII}. More precisely, we calculate
\begin{widetext}
\ba
\{\kappa \mathcal{H}[N], \mathcal{D}[\rho]\} = \mathcal{H}[\ldots] + \mathcal{D}[\ldots]  
 +\int_{\sigma}d^Dx ~ (D-1) \sqrt{q} \rho \Delta^g \left[D_a D^a - \frac{\kappa^2}{2q a(\Phi)^2} P^{\text{tf}}_{ab} P_{\text{tf}}^{ab} - \frac{1}{2} R^{(D)}  \right] N 
\ea
\end{widetext}
and conclude that $\mathcal{D}=0$ is locally a good gauge fixing if the elliptic partial differential operator
\be
	\label{eq:localgauge}
	D_a D^a - \frac{\kappa^2}{2q a(\Phi)^2} P^{\text{tf}}_{ab} P_{\text{tf}}^{ab} - \frac{1}{2} R^{(D)}  
\ee
is invertible. By a standard argument from the theory of partial differential equations \cite{GilbargEllipticPartialDifferential} and using the assumption of a compact spatial slice without boundary (boundaries and non-compact spatial slices are treated in \cite{BSTII}), it is sufficient to show that 
\be
\label{eq:invertcond}
\frac{\kappa^2}{2q a(\Phi)^2} P^{\text{tf}}_{ab} P_{\text{tf}}^{ab} + \frac{1}{2} R^{(D)} > 0 \text{.}
\ee
If we demand the dominant energy condition ($- T_{\mu} \mbox{}^{\nu} \zeta^{\mu}$ is a future causal vector for all future timelike vectors $\zeta$) and use the field equations as well as the vanishing of the constraints, it follows that
\be
	\frac{1}{2} R^{(D)} > \frac{1}{2} [K_{ab} K^{ab} - K^2] \approx \frac{\kappa^2}{2 q a(\Phi)^2}  P^{\text{tf}}_{ab} P_{\text{tf}}^{ab}  \text{,}
\ee
and thus \eqref{eq:invertcond} holds:
\ba
	 \frac{\kappa^2}{2 q ~ a(\Phi)^2} P^{\text{tf}}_{ab} P_{\text{tf}}^{ab} + \frac{1}{2} R^{(D)} > \frac{\kappa^2}{q a(\Phi)^2} P^{\text{tf}}_{ab} P_{\text{tf}}^{ab}  \geq 0 \text{.}
\ea
However, the dominant energy condition does not generally hold for the conformally coupled scalar field \cite{FordClassicalScalarFields} and we would have to impose it as additional, though physically motivated, constraint. Since the  dominant energy condition is an inequality, we expect that the dimensionality of the phase space is not reduced by its imposition. In particular, the MST black hole discussed later on is admissible. Nevertheless, it is only a sufficient condition to have \eqref{eq:invertcond} satisfied and it might possibly be relaxed.

The next step is to construct a new metric variable invariant under local conformal transformations. This new variable is built such that the unphysical conformal mode of the original metric is accounted for by the scalar field. In this way, the physical content of the original scalar field becomes part of the new metric. The unphysical degree of freedom rests in the new scalar field variable, which is reflected by the fact that its conjugate momentum is given by $\mathcal{D}$. Explicitly, this is achieved by the canonical transformation
\ba
	 \tilde{q}_{ab} &:=& e^{\frac{4}{D-1} \tilde{\phi}} q_{ab}, ~~\tilde{P}^{ab} := e^{- \frac{4}{D-1}\tilde{\phi}} P^{ab}, \nonumber \\ 
	 \tilde{\phi} &:=& \text{ln} \, \Phi, ~~\tilde{\pi}_{\tilde{\phi}} := \frac{1}{\Delta^{\Phi}} \mathcal{D}  \text{.}
\ea
Indeed, the new Poisson brackets read 
\be
	\{\tilde{q}_{ab}, \tilde{P^{cd}} \} = \delta_{(a}^c \delta_{b)}^d, ~~~ \{ \tilde{\phi}, \tilde{\pi}_{\tilde{\phi}} \} = 1 \text{.} 
\ee
Here, we restricted ourselves to $\Phi>0$, which can be interpreted as a dilaton-type field $\Phi = \exp \tilde{\phi}$. This restriction is necessary in order not to divide by zero and an according restriction on the spacetimes which we want to describe follows. 

Next, we pass to the Dirac bracket $\{\cdot,\cdot\}_{\text{DB}}$ associated with implementing $\mathcal{H}=\mathcal{D}=0$ simultaneously, which solves these constraints classically. We note that $\tilde{q}_{ab}$ and $\tilde{P}^{ab}$ are enough functions to parametrise the reduced phase space, and since $\{\tilde{q}_{ab},\mathcal{D} \} = \{\tilde{P}^{ab},\mathcal{D} \} = 0$, the non-vanishing Dirac brackets among them are
\be
	\{\tilde{q}_{ab},\tilde{P}^{cd} \}_{\text{DB}} = \{\tilde{q}_{ab},\tilde{P}^{cd} \} = \delta_{(a}^c \delta_{b)}^d \text{.}
\ee
The remaining constraint algebra simply reads
\be
\{\mathcal{H}_a[N^a], \mathcal{H}_b[M^b] \}_{\text{DB}} = \mathcal{H}_a[(\mathcal{L}_N M)^a] \text{.}
\ee
Up to the missing Hamiltonian constraint, 
this system is identical to the ADM formulation \cite{ArnowittTheDynamicsOf} of general relativity and we can thus use standard techniques from loop quantum gravity in order to quantise it.

\section*{Quantisation}

From the above ADM-type phase space, we can perform a canonical transformation to real connection variables as in \cite{AshtekarNewVariablesFor, BarberoRealAshtekarVariables}, or in all dimensions $D \geq 2$ along the lines of \cite{BTTI}. We will shortly discuss the canonical transformation in case of the Ashtekar-Barbero variables in $D=3$, see \cite{ThiemannModernCanonicalQuantum} for further details. We start from the symplectic potential 
\be
	 - (\delta \tilde{P}^{ab})  \tilde{q}_{ab} 
\ee
from the previous section. Using a tetrad defining $\tilde{q}_{ab} = \tilde{e}_a^i \tilde{e}_b^j \delta_{ij}$ introduces SO$(3)$ (or SU$(2)$) gauge symmetry and the and the symplectic potential becomes
\be
	- (\delta \tilde{P}^{ab})  \tilde{q}_{ab} = 2 \tilde{E}^{ai} \delta \tilde{K}_{ai}
\ee
with $\tilde{K}_{ai} = \tilde{K}^{bc} \tilde{q}_{ab} \tilde{e}_{ci}$, $\tilde{K}^{ab} = \frac{\kappa}{\sqrt{\tilde{q}}} (\tilde{P}^{ab}-\frac{1}{2} \tilde{P} \tilde{q}^{ab})$ and $\tilde{E}^{ai} = \tilde{e}_{b}^i \tilde{q}^{ab} \sqrt{\tilde{q}}$. Note that $\tilde{K}^{ab}$ is not just a rescaled version of the physical extrinsic curvature $K^{ab}$, but also depends on the scalar field and its momentum \eqref{eq:ADMMomentum}. The key step to derive the Ashtekar-Barbero variables is now to show that
\be
	\tilde{E}^{ai} \delta \tilde{\Gamma}_{ai} + \frac{1}{2} \epsilon^{abc} \partial_c \left[ (\delta \tilde{e}_b^j) \tilde{e}_{cj} \text{sgn}(\det (\tilde{e})) \right] =0 \text{,}
\ee
where $\tilde{\Gamma}_{ai}$ is the spin connection compatible with $\tilde{e}_{ai}$. 
The symplectic potential can thus be rewritten as
\be
	\int_\sigma d^3x \, 2\m^{(\beta)} \tilde{E}^{ai} \delta\m^{(\beta)} \tilde{A}_{ai} + \frac{1}{\beta}\int_{\partial \sigma} dS_a \,  \epsilon^{abc}   (\delta \tilde{e}_b^j) \tilde{e}_{cj} \text{sgn}(\det (\tilde{e})), \label{eq:SympPotential}
\ee
and we identified the conjugate bulk variables $\m^{(\beta)} \tilde{E}^{ai} = (1/ \beta) \tilde{E}^{ai}$ and $\m^{(\beta)} \tilde{A}_{ai} = \tilde{\Gamma}_{ai} + \beta \tilde{K}_{ai}$ with $\beta \in \mathbb{R}\backslash \{0\}$ being the Barbero-Immirzi parameter. The boundary term will be discussed later in the section on applications to black hole entropy calculations. 

A mathematically rigorous quantisation of this classical system can be accomplished by loop quantum gravity methods \cite{RovelliQuantumGravity, ThiemannModernCanonicalQuantum} and the uniqueness result on the representation \cite{LewandowskiUniquenessOfDiffeomorphism} when demanding a unitary representation of the spatial diffeomorphisms remains valid since the spatial diffeomorphism constraint still has to be quantised. The difference to loop quantum gravity is, however, that the Hamiltonian constraint has been solved already classically and the usual complications associated with its quantisation do not arise. 

The Gau{\ss} and spatial diffeomorphism constraint can be solved by standard methods \cite{ThiemannModernCanonicalQuantum}. As for spatially diffeomorphism invariant operators, in our case physical observables, we have nothing new to add to the usual treatment, see \cite{ThiemannModernCanonicalQuantum} for an exposition. Further research for a better understanding of these operators, especially graph-changing ones, is nevertheless needed.

\section*{Geometric operators}

The geometric operators of loop quantum gravity, such as the area and volume operators, can be constructed in the usual manner from the invariant connection. However, their interpretation now changes since their spectrum has to be related with the geometry based on the non-invariant metric. It follows that, morally speaking, 
\be
\hat{A}^{\text{inv}} = \Phi^2 \hat{A}^{\text{LQG}}, ~~~ \hat{V}^{\text{inv}} = \Phi^{2D/(D-1)} \hat{V}^{\text{LQG}} \text{,}
\ee
where the usual LQG operators measure the actual geometry while the invariant operators have the familiar discrete spectrum. A similar, although conceptually different, observation has been made by Ashtekar and Corichi in \cite{AshtekarNonMinimalCouplingsQuantum}. We remark that the possible occurrence of such a phenomenon has been emphasised by Dittrich and Thiemann \cite{DittrichAreTheSpectra}: The geometric operators of LQG might change their spectrum when taking into account the Hamiltonian constraint. This has to be seen in contrast to the result of \cite{GieselAQG4, DomagalaGravityQuantizedLoop}, where the spectra remain unchanged. The change in spectrum has to be attributed to the different choice of equal time hypersurfaces, i.e. $\mathcal{D} = 0$ in our case and, e.g. $\Phi - \text{const}=0$ in \cite{DomagalaGravityQuantizedLoop}, and the different resulting invariant geometric operators, which have to Poisson commute with the time function at the classical level. Further discussion on this issue is given in our companion paper \cite{BSTII}.
It is interesting to note that when using a Higgs-type potential for $\Phi$ which leads to a non-vanishing vacuum expectation value $\vev{\Phi}$, one could approximate the invariant geometric operators by the LQG geometric operators times a constant which changes the fundamental geometric scale by a factor of $1/ \vev{\Phi}$ in Planck units. While this might present a mechanism to increase the fundamental scale of LQG and make it thus more accessible to experiments, we caution the reader that such an interpretation is strongly tied to the type of foliation we are using and that the associated dynamics have to be investigated to check for consistency with current experiments, thus making further research necessary before jumping to conclusions. Also, the proposed quantisation of $\Phi$ would be very different than in the standard model, since $\Phi$ would be quantised as a part of the invariant metric instead of a usual scalar field.

Of course, at this point, these invariant operators still do not commute with the spatial diffeomorphism constraint, which could for example be achieved by tying their domain of integration to physical values of other matter fields. We leave this issue for further research.

\section*{Application to black hole entropy}

One of the major open problems in the calculation of black hole entropy in the loop quantum gravity framework is the treatment of the Hamiltonian constraint. While the constraint vanishes on the black hole horizon \cite{AshtekarIsolatedHorizonsThe} and therefore does not have to be taken into account there, it still acts on the bulk. In the entropy calculations, it is assumed that every horizon state has at least one extension into the bulk which is annihilated by the Hamiltonian constraint, a proof, however, has not been given so far. On the other hand, using the techniques developed in this paper, the problem of implementing the Hamiltonian constraint in the bulk does not even arise, since it is solved classically.  
We briefly sketch the relevant aspects of the black hole entropy calculation in our framework and comment on the outcome.
Building on the results of \cite{AshtekarNonMinimallyCoupled, AshtekarNonMinimalCouplingsQuantum, DeBenedictisPhaseSpaceAnd, DeBenedictisANoteOn}, the entropy can be calculated by counting the horizon states which are in agreement with the macroscopic properties of the black hole prescribed by the {\it invariant} area operator instead of the usual LQG area operator. 

First, we remark that several black hole solutions for general relativity conformally coupled to a scalar field exist, which avoid the no-hair theorems in $3+1$ dimensions and allow for non-trivial horizon topologies, see \cite{NadaliniThermodynamicalPropertiesOf} for an overview. In order to treat them in our framework, we first have to check if the gauge $\mathcal{D}=0$ is accessible. For simplicity, let us restrict to static, i.e. the metric and the scalar field do not depend on the time coordinate $t$, $3+1$ dimensional solutions. Choosing the $t=\text{const.}$ hypersurfaces as the leaves of our foliation, accessibility directly follows since all the momenta, and thus $\mathcal{D}$, vanish in this case. Next, we have to check if the gauge is well behaved, i.e. (\ref{eq:localgauge}) has trivial kernel (an extension to non-compact spatial slices is treated in \cite{BSTII}, where we also discuss global aspects). In the case of vanishing cosmological constant $\Lambda$, the scalar field is diverging at the horizon and we neglect this case. For $\Lambda > 0$, it was shown in \cite{BSTII} that $\mathcal{D}$ can be supplemented with an additional term to imply that (\ref{eq:localgauge}) has trivial kernel. However, this additional term would spoil the accessibility of the gauge for the $t=\text{const.}$ foliation. On the other hand, for $\Lambda < 0$, $\mathcal{D}$ may remain unaltered and triviality of the kernel still follows. The corresponding black hole solution has been found by Mart\'inez, Staforelli and Troncoso, it describes an asymptotically locally AdS black hole and admits non-trivial horizon topologies of the form $H^2 / \Gamma$, where $H^2$ is the hyperbolic plane and $\Gamma$ is a freely acting discrete subgroup of O$(2,1)$ \cite{MartinezTopologicalBlackHoles}. In the following, a spatial constant mean curvature slice of the horizon will be called $S$, its area $A$, and its Euler characteristic $\chi$. Note that $H^2 / \Gamma$ is topologically equivalent to a handle body with genus $g$ and Euler characteristic $\chi = 2(1-g)$. The case $g=1$ of a torus has to be excluded since $\chi=0$ would lead to ill-defined expressions in the following calculations. The afore mentioned black hole solution also needs a quartic self-interaction term of the scalar field, which however does not spoil the applicability of the techniques developed here, as discussed in depth in our companion paper \cite{BSTII}.

One might object that the gauge is not fixed completely in the above static spacetime, because $\mathcal{D} = 0$ can select any $t=\text{const}$ hypersurface. However, the transformation between different $t=\text{const}$ hypersurfaces is {\it not} a gauge transformation but an asymptotic symmetry and thus not a constraint which we have to fix, see also \cite{BSTII}. This can be seen by the fact that the corresponding lapse function would not vanish sufficiently fast at asymptotic infinity, thus leading to boundary terms which spoil the invertibility of (\ref{eq:localgauge}).
Still, non-trivial global problems could appear such as additional $\mathcal{D}=0$ surfaces with different shapes and $t \neq \text{const}$, corresponding to the well known problem of Gribov copies. These have to be studied in detail for the specific black hole solutions under consideration. We will neglect this potential complication in this paper.

In the following, we will briefly outline the results of the proposed state counting in the U$(1)$ framework originally introduced by Ashtekar et al. \cite{AshtekarIsolatedHorizonsThe, AshtekarMechanicsOfIsolated, AshtekarIsolatedHorizonsHamiltonian, AshtekarQuantumGeometryOf} (The SU$(2)$ framework \cite{EngleBlackHoleEntropy, EngleBlackHoleEntropyFrom} works equally well, but reference to the complete Hamiltonian treatment given in \cite{ThiemannModernCanonicalQuantum}, which we will follow closely, facilitates the discussion in this paper.) To do this, we will generalise the calculations given in chapter 15 of \cite{ThiemannModernCanonicalQuantum} to our case at the appropriate points. The definition of isolated horizons remains unchanged in presence of the conformally coupled scalar field. It is however important that the scalar field is constant on a horizon slice $S$. It is well known that due to the conformal coupling, the entropy does not obey the usual area law $S_{\text{BH}} =  A / (4G)$, but it picks up a factor of $a(\Phi)$ and the correct expression should read $S_{\text{BH}} =  a(\Phi) A / (4G)$. In the context of loop quantum gravity, this has been confirmed in \cite{AshtekarNonMinimallyCoupled, AshtekarNonMinimalCouplings} starting from a first order framework. 
Before commenting on this result and the relation to ours, we will first outline the calculation in our framework.

The important observation which allows to calculate the black hole entropy in the isolated horizon framework in loop quantum gravity is that the canonical transformation to connection variables yields a Chern-Simons symplectic potential on the boundary $S$ of $\sigma$. In order to derive this result, considers the boundary term
\be
	\frac{1}{\beta} \int_S \delta m \wedge \bar{m} \text{,}
\ee
induced by the canonical transformation in the quantisation section, where $m$ is the complex co-dyad on $S$. Note that we used $\text{sgn}(\det (\tilde{e})) = \text{sgn}(\Phi^3 \det (e)) = \text{sgn}(\Phi) \text{sgn}(\det (e)) = 1$, since $\text{sgn}(\Phi) =1$ because we restricted to $\Phi > 0$ and we can always assume $\text{sgn}(\det (e)) = 1$ classically. It can be shown that the symplectic potential 
\be
	\frac{1}{\beta} \int_S \delta m \wedge \bar{m} + \frac{1}{2c \beta} \int_S  \delta W \wedge W \label{eq:ClosedPotential}
\ee
is closed, where $W$ is the SO$(2)$ spin connection compatible with $m$ on $S$ and $c = - \frac{\chi \pi}{A}$. Thus, up to the closed form (\ref{eq:ClosedPotential}), the symplectic potential on the boundary coincides with a U$(1)$ Chern-Simons symplectic potential. In \cite{ThiemannModernCanonicalQuantum}, the calculation was restricted to $S$ being a two-sphere, thus having Euler-characteristic $\chi = 2$. The generalisation to different topologies is straight forward and has already been discussed in \cite{DeBenedictisPhaseSpaceAnd}. 

In order to generalise this calculation to our case, we have to use a conformally invariant $\tilde{m} = m \Phi$. In terms of $\tilde{m}$, the symplectic potential 
\be
	\frac{1}{\beta} \int_S \delta \tilde{m} \wedge \bar{\tilde{m}} +  \frac{ \Phi^2}{2c \beta} \int_S  \delta W \wedge W 
\ee
is closed due to $\Phi$ being constant on $S$ and being held fixed in variations. Since $W$ is a homogeneous function of degree zero of $m$ for global rescalings, it follows that $\tilde{W} = W$. The next ingredient in the calculation is the isolated horizon boundary condition 
\be
	d W = \beta c \, (*E)_j \delta^j_3
\ee
relating the bulk degrees of freedom probed by the flux $E_j$ to the field strength $dW$ of the Chern-Simons connection $W$. In the same way as above, it is replaced by 
\be
	d W = \frac{\beta c}{\Phi^2} (* \tilde{E})_j \delta^j_3 \text{.}
\ee
in our case. Instead of $c$, we obtain a rescaled constant $\tilde{c} = c / \Phi^2 = - \frac{\chi \pi}{A \Phi^2}$, which coincides with $c$ up to the fact that the area $A$ of $S$ is replaced by the invariant area $A \Phi^2$ as measured by the invariant area operator. The rest of the entropy calculation works exactly as before, just with the substitution of the invariant area instead of the area. The final result is 
\be
	S_{\text{BH}} =   \frac{A}{4 G}  \Phi^2 \frac{\beta_0}{\beta} \label{eq:BHEntropy}
\ee 
with $\beta_0$ being a constant. 
The appearance of the additional factor $\Phi^2$ in the final formula can be understood in the following way: In loop quantum gravity, the canonical variables are given by the rescaled densitised triad $\m^{(\beta)} E^a_i = (1/\beta) E^a_i $ and the conjugate SU$(2)$-connection $\m^{(\beta)} A_{ai} = \Gamma_{ai} + \beta K_{ai}$, where $K_{ai}$ is the contraction of the extrinsic curvature with the co-triad $K_{ai} = K_{ab} e^{b}_i$. In the case of a conformally coupled scalar field we are considering here, the variable conjugate to $E^a_i$ is changed from $K_b^j$ to $K'\m_b^j := a(\Phi)[K_{b}^j + (a'(\Phi)/2a(\Phi)) e_{b}^j \mathcal{L}_n \Phi]$, which follows from \eqref{eq:ADMMomentum}. Since $\Phi$ is constant on $S$, using the conformally invariant variables $\m^{(\beta)} \tilde{E}^a_{i} = (\Phi^2 / \beta) E^a_i$ and 
$\tilde{K}_{ai} = (1/\Phi^2) K'_{ai}$ in $\m^{(\beta)} \tilde{A}_{ai} = \tilde{\Gamma}_{ai}+\beta \tilde{K}_{ai}$ is exactly like using a redefined Barbero-Immirzi parameter $\tilde{\beta} = \beta / \Phi^2$. 
This parameter will consequently show up in the final expression for the entropy. Since the canonical transformation relating different choices for $\beta$ cannot be implemented on the holonomy flux algebra, the spectrum of observables and thus the black hole entropy can depend on the choice of $\beta$, or, in our case, on the effective $\tilde{\beta}$. 

It has been customary to fix the value of the Barbero-Immirzi parameter $\beta$ by this type of entropy calculation to $\beta_0$ in order to obtain the familiar area law. The validity of this procedure has been checked for many different types of black holes and matter contents of the theory by showing that $\beta_0$ is always the same number. 
However, we see a problem with this procedure for our specific choice of black hole: 
the value of $\Phi$ on S enters the entropy formula not in the wanted expression $a(\Phi)$. Since $\Phi$ depends on the mass of the black hole \cite{MartinezTopologicalBlackHoles}, fixing $\beta$ is not an option. 
This wrong dependency on $\Phi$ is closely connected to the choice of effective Barbero-Immirzi parameter as discussed below. It can be evaded by a different choice of variables which are not conformally invariant as explained at the end of this section. First however, we will focus on a different route.

A possible way to solve this problem has been recently proposed in \cite{FroddenBlackHoleEntropy}, 
where the authors observe that an analytic continuation of the formula for the dimension of the state space of the Chern-Simons theory compatible with the macroscopic area $A$ to the complex value $\beta = \pm i$,  yields the area law $S_{\text{BH}} = A / (4G) + \text{corrections}$. 
The choice $\beta = \pm  i$ plays a distinguished role since the corresponding Ashtekar connection is the pullback of the (anti) self-dual spacetime spin connection. However, one has to keep in mind that the quantum theory for $\beta = \pm i$ is ill-defined and the derivation is only formal.

Still applying the method of \cite{FroddenBlackHoleEntropy} to the case at hand, we first have to clarify which value of $\beta$ needs to be chosen in order that the corresponding connection plays the same distinguished role. A first order action for the conformally coupled scalar field using (anti) self-dual connections and the canonical analysis thereof was given in \cite{CapovillaNonminimallyCoupledScalar}. The canonical variables, adapted to our notation, turn out to be $\hat{E}^{a}_i := \mp i a(\Phi) E^{a}_i$ and $\hat{A}_{bj} := \hat{\Gamma}_{bj} \pm i \hat{K}_{bj}$, where $\hat{\Gamma}_{bj}$ is compatible with $\hat{E}^a_i$ and $\hat{K}_{bj} := (1/a(\Phi)) K'_{bj}$ (the scalar field momentum $\pi_{\Phi}$ is modified accordingly but its form is not relevant in what follows). $\hat{A}_{bj}$ is the pullback of the on-shell self-dual spacetime connection (i.e. if the equations of motion hold). Comparison with the conformally invariant variables \eqref{eq:SympPotential} reveals that in the presence of a conformally coupled scalar field, one has to analytically continue $\beta$ to $\pm i \Phi^2 / a(\Phi)$. This obviously transforms $(1/\beta) \tilde{E}^{ai}$ and $\beta \tilde{K}_{bj}$ into $\hat{E}^{ai}$ and $\pm i \hat{K}_{bj}$. Furthermore, in the spatial bulk, setting this value for $\beta$ corresponds to a local conformal transformation, and to implement it canonically, we have to change the connection part $\tilde{\Gamma}_{bj}$ in $\tilde{A}_{ai}$ to $\hat{\Gamma}_{bj}$. In total, the analytical continuation $\beta \rightarrow  \pm i \Phi^2 / a(\Phi)$ transforms $\tilde{A}_{ai}$ into the (anti) self-dual connection $\hat{A}_{ai}$.

Going through the calculations of \cite{FroddenBlackHoleEntropy} for the case at hand, the factor $\Phi^2$ appearing in \eqref{eq:BHEntropy} is canceled by a similar factor from the invariant area operator, giving the final result $S_{\text{BH}} =  a(\Phi) A / (4G)$, i.e. the needed factor $a(\Phi)$ is recovered. 
We emphasise that we currently don't understand why this method yields the correct result or what the deeper implications it might have. For now, we consider it as an ad-hoc procedure which yields the correct result in a case where the usual prescription of setting $\beta = \beta_0$ fails.

A further comment is in order: At first sight, (\ref{eq:BHEntropy}) seems to contradict the results of \cite{AshtekarNonMinimallyCoupled, AshtekarNonMinimalCouplings}, where an entropy proportional to $a(\Phi)$ was obtained by fixing $\beta = \beta_0$. However, in that work, one started from a first order framework which naturally leads to the conjugate variables $\hat{K}_{ai}$ and $\pm i \hat{E}^{ai}$. As we have seen before, they are related with the conformally invariant variables we use by a rescaling with $\Phi^2/a(\Phi)$ (and a corresponding modification in the momentum $\pi_{\Phi}$ to make the transformation canonical).
While this choice of variables is classically equally valid and leads to an additional factor of $a(\Phi)/\Phi^2$ in the entropy due to a further modified effective Barbero-Immirzi parameter, it is not conformally invariant and the state counting cannot be based on physical states by the methods of this paper. We remark that this is not meant as a criticism of this choice of variables, it just means that these variables are not well suited for the constant mean curvature gauge fixing used in this paper. 

The standard loop quantum gravity entropy calculation thus generalise rather straightforwardly to the proposed version of loop quantum gravity on a constant mean curvature slice. However, it turns out that fixing $\beta = \beta_0$ in order to obtain the well known area law $S_{\text{BH}} = a(\Phi) A / (4G)$ does not work in this case since the value of the scalar field at the horizon enters not as expected. The recently proposed method of analytically continuing $\beta$ to obtain a self-dual connection however solves this problems, which suggests to pursue the line of research started in \cite{FroddenBlackHoleEntropy} further.

 \section*{Application to loop quantum cosmology}

The area gap of full loop quantum gravity is a key input into the loop quantum cosmology framework, see e.g. \cite{AshtekarLoopQuantumCosmology}, especially in the improved dynamics given by the $\bar{\mu}$ scheme \cite{AshtekarQuantumNatureOf}, since it prescribes the minimal area around which the field strength in the Hamiltonian constraint is approximated via holonomies. On the other hand, the results of this paper show that one can choose a time function in full loop quantum gravity which is compatible with spatial homogeneity and isotropy, e.g. $K \propto \dot{a}/a$ for flat Friedmann-Lemaitre-Robertson-Walker models, and yields a field dependent area gap, which affects the Hamiltonian constraint based on this choice of variables in a non-trivial way. 

From a phenomenological point of view, we expect that comparison with experiment would yield a lower bound on the magnitude of the scalar field (modulo the Barbero-Immirzi parameter), since otherwise the fundamental geometric scale would be too large and in conflict with cosmological observations. However, loop quantum cosmology based on the presented variables is also interesting from a technical point of view, since it requires us to polymerise not only the gravitational variables, but also the scalar field. Otherwise, the difference equation resulting from the Hamiltonian constraint, which now has a field dependent step size, would be in conflict with the technical nature of the almost periodic functions used in the construction of the loop quantum cosmology Hilbert space, since these are excited only at a finite number of points.

\section*{Concluding remarks}

\begin{itemize}

\item We underline that the original idea of trading the Hamiltonian constraint for a local conformal invariance originated in shape dynamics \cite{GomesEinsteinGravityAs}. The main new input in our formalism is that a conformally coupled scalar field allows for a non-trivial conformal scaling and thus for the construction of an invariant metric from which quantisation can start. Also, we do {\it not} have a global Hamiltonian as in \cite{GomesEinsteinGravityAs} since we are not restricting to volume preserving conformal transformations. 

\item An extension to standard model matter, a cosmological constant, and non-compact spatial slices is discussed in our companion paper \cite{BSTII}.

\item Since dilaton fields are naturally appearing in supergravity, we plan to investigate an extension of our framework to this setting. Here, it will be interesting to check what extent of supersymmetry is compatible with the $\mathcal{D}=0$ gauge fixing or how one could also gauge fix the supersymmetry constraint. 

\end{itemize}

\begin{acknowledgments} 
NB and AT thank the German National Merit Foundation for financial support. AS thanks the Evangelisches Studienwerk Villigst for financial support. NB was supported in part by the NSF Grant PHY-1205388 and the Eberly research funds of The Pennsylvania State University. We thank Abhay Ashtekar, Kristina Giesel, Henrique Gomes, Sean Gryb, Yasha Neiman, Tim Koslowski, and Thomas Thiemann for helpful and stimulating discussions. Also, we thank Kristina Giesel and Thomas Thiemann for carefully reading a draft of this paper and proposing numerous improvements. The idea for this work was born at the ESF-funded Quantum Gravity Colloqium 6, where Sean Gryb urged us to investigate the possibility of a connection formulation of shape dynamics with nice transformation properties under local conformal transformations.

\end{acknowledgments}

\end{document}